\documentclass[11pt, a4paper]{article} 
\usepackage{amsmath}
\usepackage{amssymb}
\usepackage{mathrsfs}
\usepackage{graphicx}
\usepackage{hyperref}
\hypersetup{
    colorlinks=true,  
    linkcolor=blue,   
    citecolor=blue
}
\usepackage{xcolor}

\newcommand{\R}{\mathbb{R}}

\begin{document}
\title{Completing correlation matrices}

\author{Olaf Dreyer\thanks{OD Consulting, email: olaf.dreyer@gmail.com}
\and Horst K\"ohler \thanks{Commerzbank AG}
\and Thomas Streuer \thanks{RIVACON GmbH} }

\date{\today}

\maketitle

\begin{abstract}
	We describe a way to complete a correlation matrix that is not fully specified. Such matrices often arise in financial applications when the number of stochastic variables becomes large or when several smaller models are combined in a larger model. We argue that the proper completion to consider is the matrix that maximizes the entropy of the distribution described by the matrix. We then give a way to construct this matrix starting from the graph associated with the incomplete matrix. If this graph is chordal our construction will result in a proper correlation matrix. We give a detailed description of the construction for a cross-currency model with six stochastic variables and describe extensions to larger models involving more currencies.
\end{abstract}



\section{Introduction}
Modern financial models are driven by a large number of stochastic variables. These normally distributed stochastic variables are usually correlated which creates the need to determine the coefficients of the underlying correlation matrix. Since the number of correlation coefficients increases quadratically with the number of stochastic variables, this number can become very large. While it would be desirable to calibrate all these coefficients this is in practice not possible because such a calibration procedure would be too slow, too unstable, or would require calibration instruments for which quotes can not be obtained. The issue of stability is especially important in a risk neutral setup where daily calibrations might display undesired fluctuations if the calibration is not stable. Hybrid models are another source of undetermined correlation coefficients. Two calibrated models might be parts of a larger model. While the coefficients for the individual models have been determined in separate calibrations, the coefficients that mix variables from the two different models are undetermined. In a large model, with many smaller constituent models, the number of such coefficients can constitute the bulk of the matrix. 

Faced with this situation, it is tempting to set the missing coefficients to zero. The idea here is that any other choice seems to add arbitrary information. This approach has several shortcomings though. The most important of which is that one might end up with a matrix that is not positive semidefinite and thus not a valid correlation matrix. To obtain a valid matrix, further steps are required. One possibility is to use the fact that the set of positive definite correlation matrices is an open set that contains the unit matrix. It is thus always possible to scale the off-diagonal elements of the matrix with a common scale factor $\lambda<1$ and obtain a positive definite matrix. Another possibility is to choose the nearest correlation matrix in a suitable topology (see e.g \cite{higham}). These additional steps might change all the coefficients of the matrix, including those coefficients, that have been obtained in previous calibrations. They also rely on somewhat arbitrary further assumptions like a particular notion of distance on the space of matrices. 

The approach presented here does not suffer from these shortcomings. We start with a particular criterium that singles out one particular completion. The criterium is distinguished by the fact that it puts the least amount of additional constraints on the distribution described by the completed matrix. As it turns out, this is not the matrix that contains zeroes in the places of unknown coefficients. Rather, it is the inverse of the matrix that contains zeroes in those places that were undetermined before. This criterium removes the arbitrary nature of other constructions. 

We then show how to construct this completion. The construction starts by looking at the graph associated with the incomplete matrix. The vertices of the graph are the rows (or columns) of the matrix. Since we are looking at a correlation matrix, the vertices correspond to the stochastic variables of the model. Two vertices are connected by an edge in the graph if the corresponding correlation coefficient in the matrix is known. It is here that we encounter the one condition that needs to be satisfied for the construction to work: The graph that we have just constructed needs to be chordal. A graph is chordal if every loop in the graph that has more than three edges contains a chord, i.e. an edge that connects two non-adjacent vertices in the loop. If this condition is satisfied we can follow a construction first introduced by \cite{grone} and refined by \cite{smith} to complete the matrix. Every step of the construction adds new vertices to the edge which in turn correspond to entries in the correlation matrix. The matrix obtained in this way will automatically be a positive-semidefinite so that no further steps need to be taken to obtain a valid correlation matrix. 

The paper now consists of three parts. In the next section we discuss the completion criterium. The matrix can be characterized in three different ways and we present them here. We then discuss the construction of the completion in section \ref{sec.procedure}. We do this by following the concrete example of a cross-currency model consisting of two interest rate models that are linked by one exchange rate. The two interest rate models and the model for the exchange rate are assumed to be stochastic volatility models so that we end up with a total of six stochastic variables. We will use this model to describe all the required steps. These steps can easily be generalized to more complicated models. The $N$-currency model with $(N-1)$ foreign exchange rates is one such model. It is described in section \ref{sec.several}. 

\section{The completion criterium}\label{sec.criterium}
Given a correlation matrix with unknown coefficients what is the right way to complete it? To motivate the answer we need to introduce some notation. Let 
\begin{equation}
	X = ( x_1, \ldots, x_k )^T\in \R^k
\end{equation}
be a $k$-dimensional random variable with a density function given by a $k$-dimensional normal distribution
\begin{align}
	f(X) & = N_k[\mu, H_X]( X ) \\
	& = (2\pi)^{-k/2}\det(H_X)^{-1/2}\exp\left(-\frac{1}{2}(X -\mu)^T H_X^{-1} (X-\mu)\right),
\end{align}
with mean $\mu$ and covariance matrix $H_X$. A crucial property of the normal distribution is that if we condition on some of the $x_i$, $i=1, \dots, k$, we again obtain a normal distribution. Let us assume that we want to condition on the last $l$ components of $X$:
\begin{equation}
	Z = (x_{k-l+1}, \ldots, x_k)^T \in \R^l.
\end{equation}
We partition the matrix $H$ in a way that reflects this partition of $X$:
\begin{equation}
	H_X = \begin{pmatrix}
		A & B \\
		B^T & C
	\end{pmatrix},
\end{equation}
with $A\in M_{k-l}$, $C\in M_l$, and $B\in M_{k-l,l}$. Let $\bar X$ be the first $k-l$ components of $X$:
\begin{equation}
	\bar X = (x_1, \ldots, x_{k-l})^T\in \R^{k-l}
\end{equation}
If we assume that $X$ has zero mean (i.e. $\mu=0$), the density function for $X$ conditioned on $Z$ is then given by
\begin{equation}
	f(\bar X \vert Z ) = N_{k-l}[BC^{-1}Z, H_X/C ](\bar X),
\end{equation}
(see \cite[chapter 8]{rao} and \cite{cottle}) where $H_X/C$ is the Schur complement of $C$ in $H_X$:
\begin{equation}\label{eqn.hxc}
	H_X/C = A - B C^{-1}B^T
\end{equation}
(Emilie Haynsworth introduced the name and highlighted its usefulness in \cite{hayns}. Schur originally made use of it in \cite{schur}. For an overview of the properties of the Schur complement see \cite{horn}.) We see that we again obtain a normal distribution. We will meet the Schur complement again soon.

Now let us look at a second set of random variables 
\begin{equation}
	Y = (x_{k-l+1}, \ldots, x_k, x_{k+1}, \ldots, x_n)^T \in \R^{n-k+l},
\end{equation}
that has an overlap $Z$ with the random variables $X$. Let the density function for $Y$ be given by
\begin{equation}
	f(Y) = N_{n-k+l}[0,H_Y] ( Y),
\end{equation}
with
\begin{equation}
	H_Y = \begin{pmatrix}
		C & D \\
		D^T & E
	\end{pmatrix}.
\end{equation}
Note that the submatrix $C$ is shared with $H_X$. In our context the matrix $C$ might have been obtained by a previous calibration of a model that is part of both $X$ and $Y$. If we were to condition on $Z$ we would find
\begin{equation}
	f(\bar Y\vert Z) = N_{n-k-l}[ D^T C^{-1}Z,H_Y/C](\bar Y), 
\end{equation}
with
\begin{equation}\label{eqn.hyc}
	H_Y/C = E - D^TC^{-1}D
\end{equation}
and $\bar Y$ is the part of $Y$ that is not $Z$. If we now want to describe $X$ and $Y$ together we are looking at the matrix
\begin{equation}
	H = \begin{pmatrix}
		A & B & W \\
		B^T & C & D \\
		W^T & D^T & E
	\end{pmatrix},
\end{equation}
with a yet to be determined matrix $W\in M_{k-l,n-l}$. Because $X$ and $Y$ share the random variables in $Z$ we can not make them independent. The next best thing that we can do is to demand that if we fix $Z$ the remaining parts of $X$ and $Y$ are independent. The combined density for $X$ and $Y$ when we condition on $Z$ is given by
\begin{equation}
	f( \bar X,\bar Y\vert Z) = N_{n-l}[ (B^T,D)^TC^{-1}Z, H/C](\bar X, \bar Y),
\end{equation}
with
\begin{align}
	H/C & = \begin{pmatrix}
		A - B C^{-1}B^T & W - B C^{-1} D \\
		W^T - D^TC^{-1}B^T & E - D^TC^{-1}D
	\end{pmatrix}\\
  & = \begin{pmatrix}
		H_X/C & W - B C^{-1} D \\
		W^T - D^TC^{-1}B^T & H_Y/C
	\end{pmatrix}.
\end{align}
We see that if we set
\begin{equation}\label{eqn.choice}
	W = BC^{-1}D
\end{equation}
the above matrix becomes block diagonal and the conditional densities obey
\begin{equation}\label{eqn.independent}
	f(\bar X,\bar Y\vert Z ) = f(\bar X\vert Z)f(\bar Y\vert Z)
\end{equation}
because the blocks in the diagonal are $H_X/C$ and $H_Y/C$ from equations (\ref{eqn.hxc}) and (\ref{eqn.hyc}). Given that $Z$ is part of both $X$ and $Y$ this is as much independence as we can ask for. The way to choose the matrix $W$ and complete $H$ is to demand that $X$ and $Y$, when conditioned on their shared part $Z$, are independent.

This choice of $W$ can be characterized in two more ways (see \cite{smith}). Banachiewicz showed that the inverse of a matrix can be expressed using the Schur complement (see \cite{horn}). Since the Schur complement $H/C$ is block diagonal the inverse of $H$ is given by
\begin{equation}
	H^{-1} = \begin{pmatrix}
		(H_X/C)^{-1} & -(H_X/C)^{-1} B C^{-1} & 0 \\
		- C^{-1} B^T (H_X/C)^{-1} & \Xi & - C^{-1} D (H_Y/C)^{-1} \\
		0 & - (H_Y/C)^{-1} D^T C^{-1} & (H_Y/C)^{-1} 
	\end{pmatrix},
\end{equation}
with
\begin{equation}
\Xi = C^{-1} + C^{-1} B^T (H_X/C)^{-1} B C^{-1} + C^{-1} D (H_Y/C)^{-1} D^T C^{-1}.
\end{equation}
We see that $H^{-1}$ has zeroes in the places where $W$ sits in $H$. The $W$ from equation (\ref{eqn.choice}) is the unique choice with that property.  

When looking at the determinant of $H$ we find another characterization of $W$. Using the Schur complement again, we find
\begin{align}
	\det H & = \det C \; \det H/C \\
	& \le \det C \; \det H_X/C \; \det H_Y/C,\label{eqn.fischer}
\end{align}
where we have equality in equation (\ref{eqn.fischer}) if and only if the off-diagonal blocks vanish, i.e. if and only if $W$ is given by equation (\ref{eqn.choice}). This choice of $W$ is the unique choice that maximizes the determinant of $H$.

The entropy $S$ of the normal distribution $N_n[\mu,H]$ is given by
\begin{equation}
	S = \frac{1}{2}\log \det H + \text{\ const.}
\end{equation} 
Because we are maximizing the determinant of $H$ we are also maximizing the entropy $S$ of the distribution described by $H$. The $H$ chosen by the $W$ from equation (\ref{eqn.choice}) is the one that puts the fewest constraints on the distribution of $X$ and $Y$. 

\section{The completion procedure}\label{sec.procedure}
In the last section we have completed the matrix $H$ in the very simple case where there are two known matrices $H_X$ and $H_Y$ with just one shared part $C$. The general case is usually more complicated. Building on the result in \cite{grone} Smith \cite{smith} (see also \cite{georgescu}) showed how to solve the general problem by reducing it to a repeated application of the result from the previous section. We will discuss the required steps in this section by looking at a particular example: the cross currency model.

\begin{figure}[hbt]
  \begin{center}
  	\includegraphics[width=10cm]{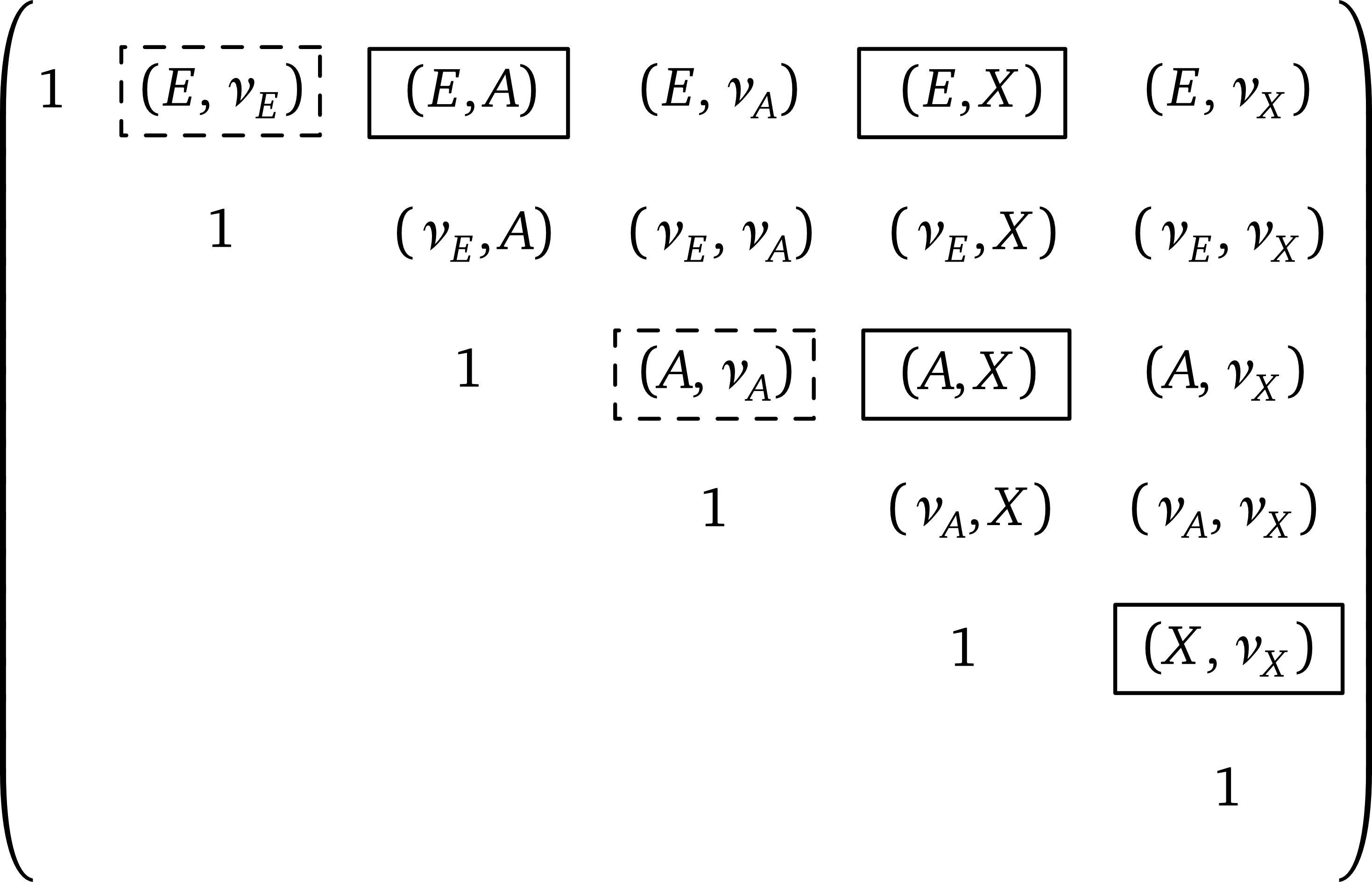}
  \end{center}
  \caption{The correlation matrix for the cross currency model. The coefficients $(E, \nu_E)$ and $(A, \nu_A)$ (the boxes with the dashed lines) are determined in separate interest rate calibrations. The coefficients $(E, X)$, $(E, A)$, $(A, X)$, and $(X, \nu_X)$ (the boxes with the solid lines) will be determined in the cross currency calibration itself. The remaining coefficients are the result of the completion procedure.}\label{fig.xccymodel}
\end{figure}

Our model will consist of two interest rate models; one for the domestic currency $E$ and one for the foreign currency $A$. These two currencies are linked by an exchange rate $X$. We will model the two interest rates and the exchange rate with stochastic volatility models so that we have a total of six stochastic variables:
\begin{equation}\label{eqn.stochvar}
	E, \nu_E, A, \nu_A, X, \nu_X
\end{equation}
We will denote the corresponding correlation coefficients by 
\begin{equation}
	(a,b),
\end{equation}
where $a$ and $b$ are one of the stochastic variables from equation (\ref{eqn.stochvar}). In total there are 15 different correlation coefficients in the correlation matrix $H$ of our model. We assume that the coefficients 
\begin{equation}
	(E,\nu_E)\text{\ and\ }(A,\nu_A)
\end{equation}
have already been determined in separate calibrations of the interest rate models. That leaves us with 13 free coefficients in $H$. This number of coefficients is too large for a stable calibration. Instead we will only calibrate the coefficients
\begin{equation}
	(E,A), (E,X), (A,X),(X,\nu_X),
\end{equation}
and determine the remaining nine coefficients using the completion procedure described here (see figure \ref{fig.xccymodel}).

\begin{figure}[hbt]
  \begin{center}
  	\includegraphics[width=10cm]{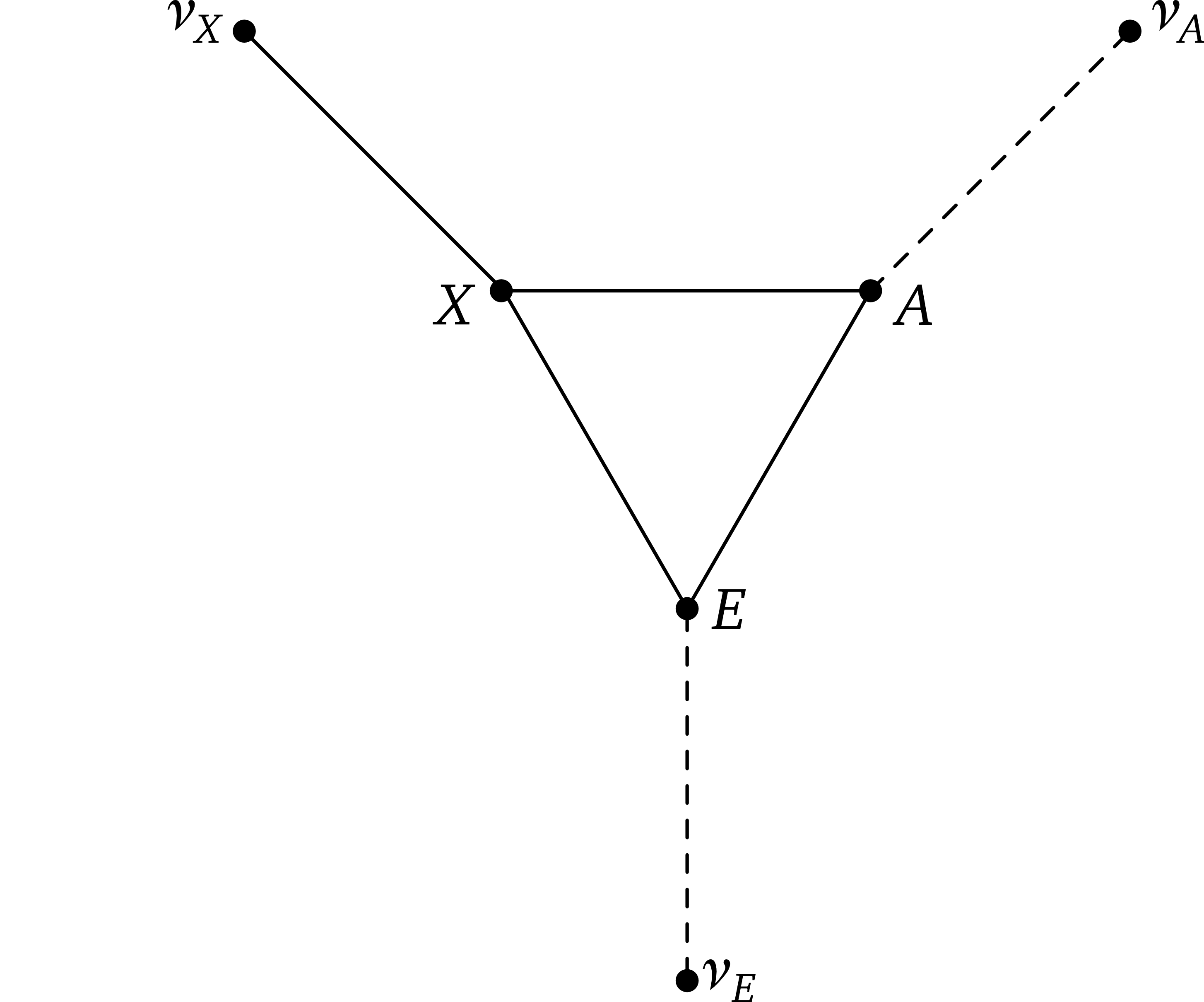}
  \end{center}
  \caption{The initial graph $\gamma$ for the cross currency calibration. The coefficients $(E,\nu_E)$ and $(A,\nu_A)$ (dashed edges) are determined in separate interest rate calibrations. The other edges are determined in the cross currency calibration itself.}\label{fig.fxCalibrationGraph}
\end{figure}

The first step to complete the matrix $H$ is to determine the graph $\gamma$ corresponding to $H$. The vertices of the graph are given by the row (or column) indices of $H$. In our case these are identified with the stochastic variables in equation (\ref{eqn.stochvar}). We then obtain the graph for $H$ by connecting the vertices for which we have coefficients. In our example we obtain the graph in figure \ref{fig.fxCalibrationGraph}.

Now that we have the graph we need to determine if the graph is chordal. A graph is chordal if every loop of length four or more has a chord (i.e. an edge connecting two non-consecutive vertices in the loop; see e.g. \cite{golumbic}). In our example this is easy to check. The largest loop is the central loop consisting of the vertices $E$, $A$, and $X$. This loop has just three elements and thus does not require a chord. The procedure that we are describing in this section requires a chordal graph.

\begin{figure}[hbt]
  \begin{center}
  	\includegraphics[width=10cm]{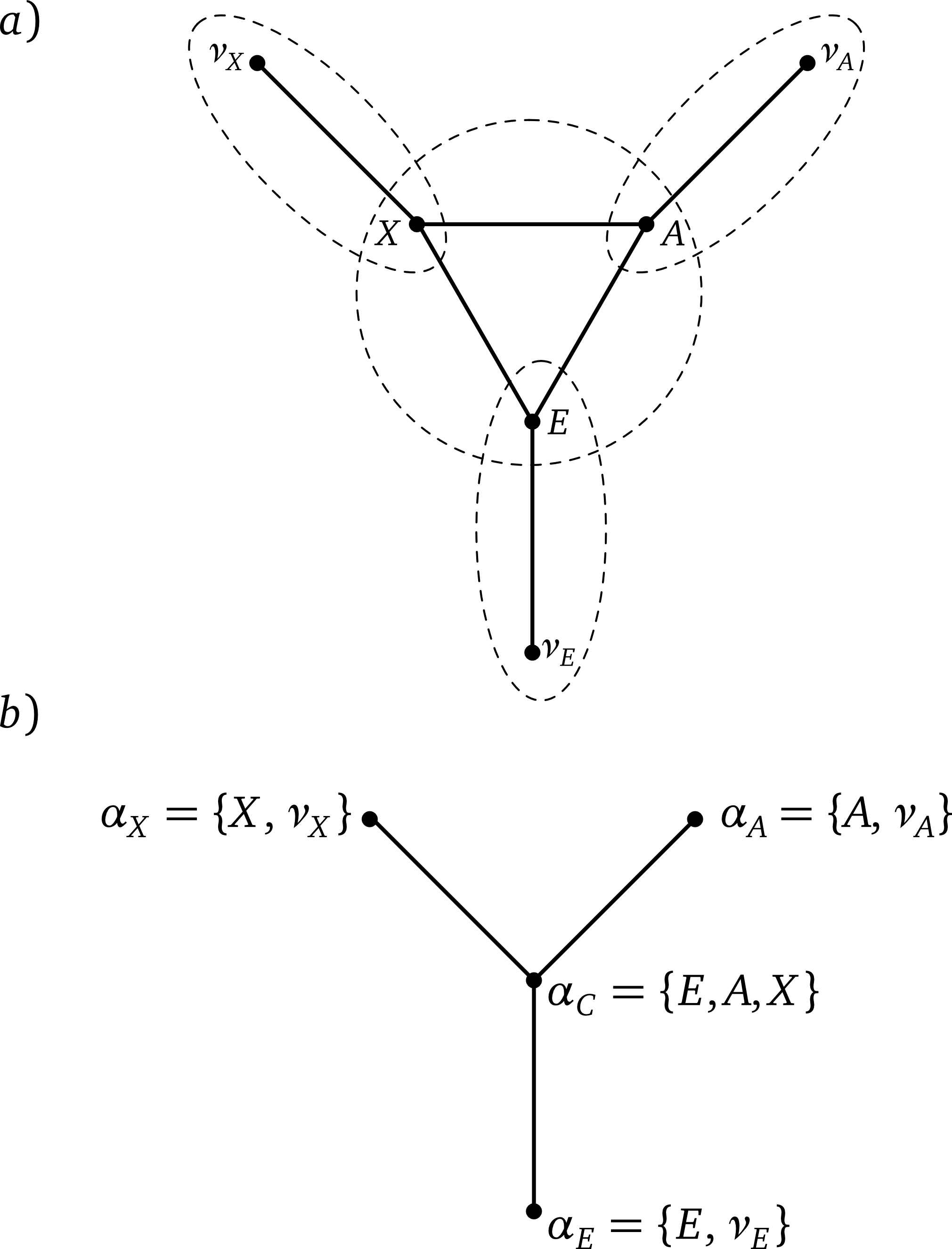}
  \end{center}
  \caption{a) The graph for the cross currency has four cliques. b) These cliques are the vertices of a new graph $\Gamma$. An edge in this new graph connects two cliques that have a non-empty intersection.}\label{fig.cliques}
\end{figure}

The next step is to identify the cliques in our graph. A clique is a maximal set of vertices that are all connected to each other (the subgraph induced by the set of vertices is complete; again, see \cite{golumbic}). In our example there are four such cliques given by the central loop and the three arms of the graph:
\begin{align}
	\alpha_C & = \{E,A,X\}\\
	\alpha_E & = \{E,\nu_E\}\\
	\alpha_A & = \{A,\nu_A\}\\
	\alpha_X & = \{X,\nu_X\}
\end{align}  
Note that the cliques have non-trivial intersections. We can use these intersections to construct a new graph $\Gamma$. The vertices of this new graph are given by the cliques $\alpha_C, \ldots, \alpha_X$ and we connect two cliques if their intersection is non-empty (see figure \ref{fig.cliques}). 

Because the graph $\gamma$ for the matrix $H$ is chordal, there exists a clique tree, i.e.  a spanning tree for the graph $\Gamma$ that possesses the intersection property. A spanning tree for $\Gamma$ is a tree that has the same vertices as the graph itself. It possesses the intersection property if for any two vertices $\alpha_1$ and $\alpha_2$ the intersection $\alpha_1\cap\alpha_2$ is contained in all the vertices on the unique path in the tree that connects $\alpha_1$ to $\alpha_2$ (see \cite{blair} for the different characterizations of a chordal graph). In our example the graph $\Gamma$ is itself a tree. It is also easy to check that it possesses the intersection property. In the next section we will see an example where the intersection property is not trivially fulfilled. We will choose $\alpha_E = \{E, \nu_E\}$ as the root of the tree. 

\begin{figure}[hbt]
  \begin{center}
  	\includegraphics[width=10cm]{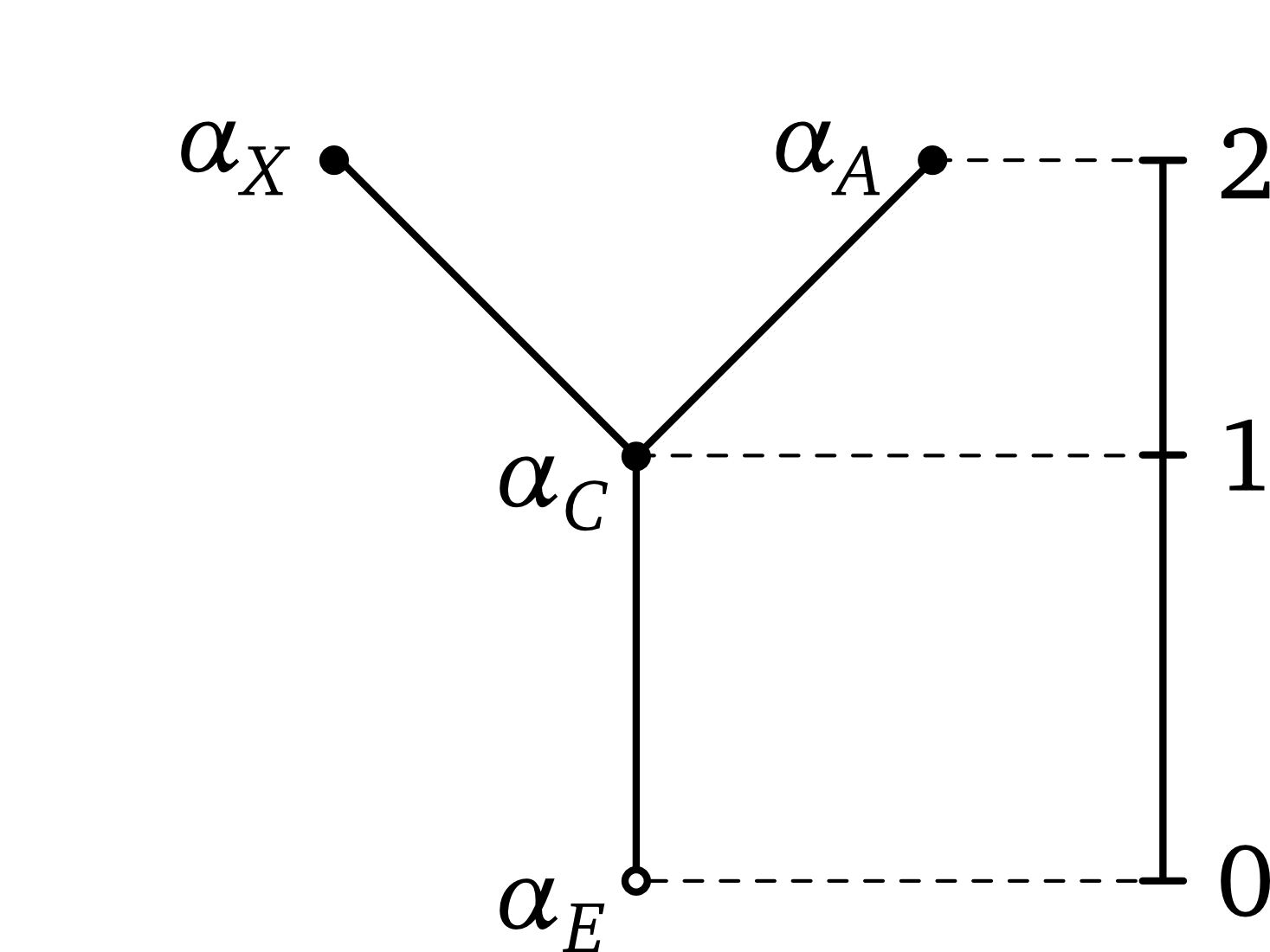}
  \end{center}
  \caption{The clique tree for the graph $\gamma$. The numbers on the right denote the height in the tree if we choose $\alpha_E$ to be the root of the tree. The completion procedure proceeds from the top of the tree to the bottom. We complete the matrix by repeatedly using equation (\ref{eqn.choice}) from the previous section.}\label{fig.cliqueTree}
\end{figure}

\begin{figure}[hbt]
  \begin{center}
  	\includegraphics[width=10cm]{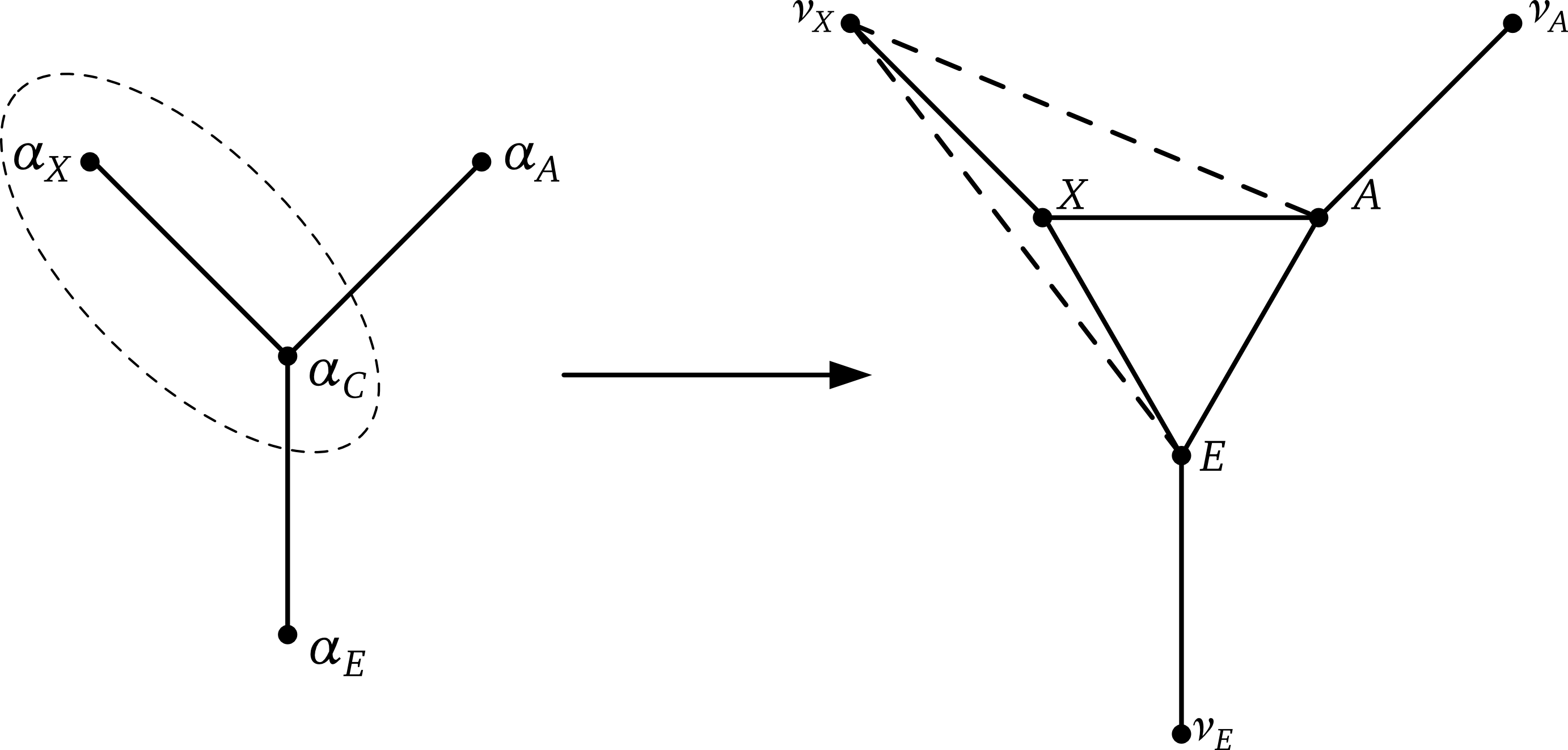}
  \end{center}
  \caption{We start the completion of the matrix $H$ with the cliques of height two. Applying equation (\ref{eqn.choice}) to the matrices given by the cliques $\alpha_X$ and $\alpha_C$ adds the dashed lines to the graph $\gamma$.}\label{fig.firstStep}
\end{figure}

We now start completing the matrix $H$ by repeatedly applying equation (\ref{eqn.choice}) from the previous section. We start with the cliques on the top of the tree and work our way down. We start with the clique $\alpha_X=\{X, \nu_X\}$. It intersects the clique $\alpha_C$ in the element $X$. This leads us to make the following identifications:
\begin{align}
	H_X & = \begin{pmatrix}
		1 & (X, \nu_X) \\
		& 1
	\end{pmatrix} \\
	H_Y & = \begin{pmatrix}
		1 & (E,X) & (A,X) \\
		& 1 & (E,A) \\
		& & 1 
	\end{pmatrix}
\end{align}
Note that the common matrix $C$ is just given by 1. Equation (\ref{eqn.choice}) now gives:
\begin{align}
	W & = ( (E,\nu_X), (A,\nu_X) ) \\
	& = BC^{-1}D \\
	& = (X, \nu_X) ( (E,X), (A,X) ) ) \\
	& =  ( (X, \nu_X)(E,X), (X, \nu_X)(A,X) ) ), 
\end{align}
or 
\begin{align}
	(E,\nu_X) & = (X, \nu_X)(E,X) \\
	(A,\nu_X) & = (X, \nu_X)(A,X).
\end{align}
Figure \ref{fig.firstStep} shows the first completion step in the graph $\gamma$. 

\begin{figure}[hbt]
  \begin{center}
  	\includegraphics[width=10cm]{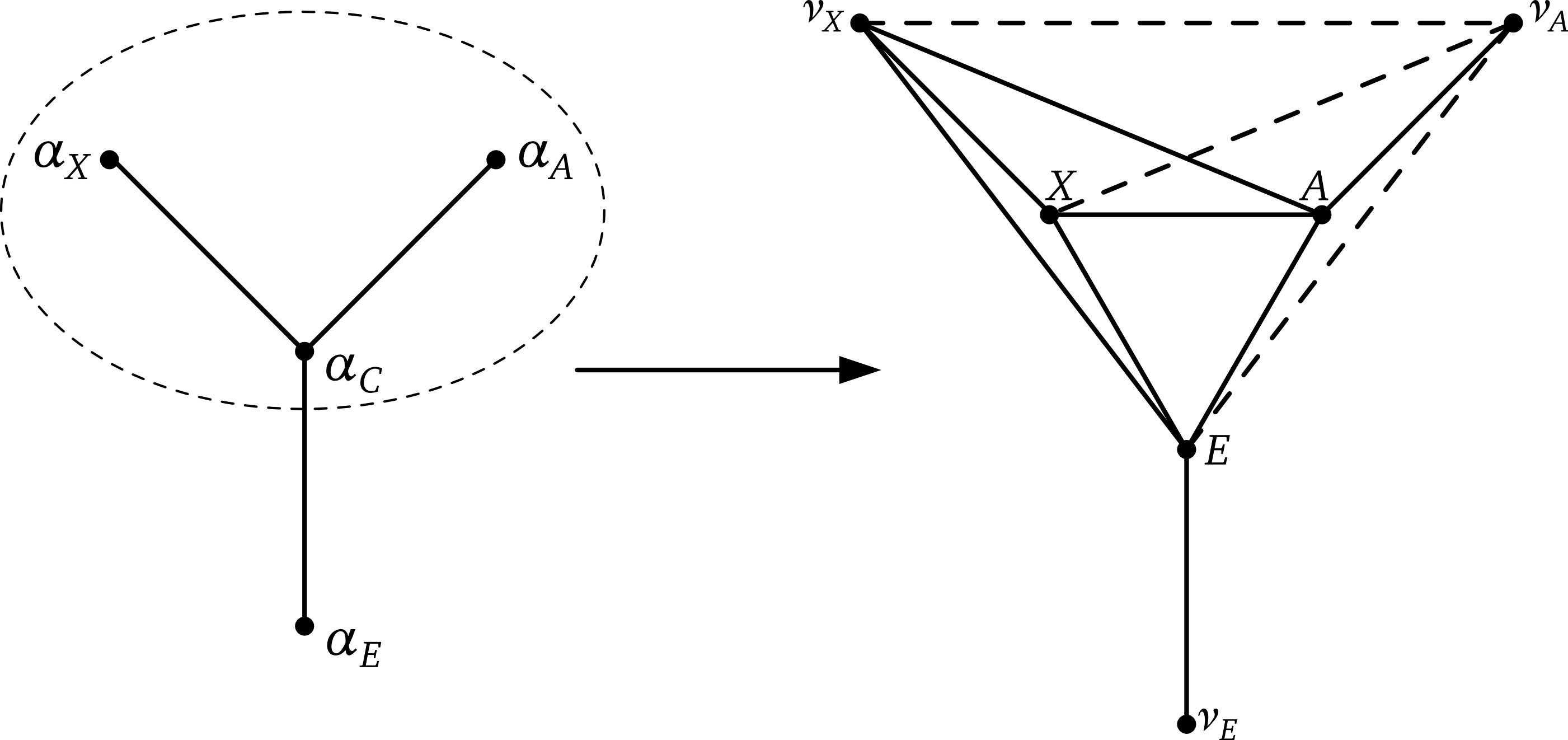}
  \end{center}
  \caption{In the second step we combine the clique from the first step with the remaining clique of height two in the tree. This step adds the three edges $(\nu_X,\nu_A)$, $(X,\nu_A)$, and $(E,\nu_A)$.}\label{fig.secondStep}
\end{figure}

We have one more clique with height two in the clique tree for $\gamma$. In the next step we take the clique $\alpha_A$ and combine it with the newly created clique from the last step. This time we make the identifications
\begin{align}
	H_X & = \begin{pmatrix}
		1 & (A, \nu_A) \\
		& 1
	\end{pmatrix} \\
	H_Y & = \begin{pmatrix}
		1 & (A,E) & (A,X) & (A,\nu_X) \\
		& 1 & (E,X) & (E,\nu_X) \\
		& & 1 & (X,\nu_X) \\
		& & & 1 
	\end{pmatrix}
\end{align}
We apply equation (\ref{eqn.choice}) again to obtain:
\begin{align}
	W & = ( (E,\nu_A), (X,\nu_A), (\nu_X,\nu_A) ) \\
	& = BC^{-1}D \\
	& = (A,\nu_A) ( (E,A), (A,X), (A,\nu_X) ),
\end{align}
or
\begin{align}
	(E,\nu_A) & = (A,\nu_A) (E,A) \\
	(X,\nu_A) & = (A,\nu_A)(A,X)\\
	(\nu_X,\nu_A) & = (A,\nu_A) (A,\nu_X) \\
	& = (A,\nu_A)(X, \nu_X)(A,X)
\end{align}
Figure \ref{fig.secondStep} shows the result of this step in the graph $\gamma$. 

There are no more cliques of height two left. Looking at the graph in figure \ref{fig.secondStep} we see that only the edges connecting $\nu_E$ to the large clique that is the result of the last step are missing. These missing edges will be created in the last step. We make the following identifications:
\begin{align}
		H_X & = \begin{pmatrix}
		1 & (E, \nu_E) \\
		& 1
	\end{pmatrix} \\
	H_Y & = \begin{pmatrix}
		1 & (E,A) & (E,\nu_A) & (E,X) & (E,\nu_X) \\
		& 1 &  (A,\nu_A) & (A,X) & (A,\nu_X) \\
		& & 1 & (\nu_A,X) & (\nu_A,\nu_X) \\
		& & & 1 & (X,\nu_X) \\
		& & & & 1 
	\end{pmatrix}
\end{align}
Applying equation (\ref{eqn.choice}) one more time gives:
\begin{align}
	W &= ( (\nu_E,A),(\nu_E,\nu_A),(\nu_E,X),(\nu_E,\nu_X) ) \\ 
	& = BC^{-1}D \\
	& = (E, \nu_E) ( (E,A),(E,\nu_A),(E,X),(E,\nu_X) ), 
\end{align}
or
\begin{align}
	(\nu_E,A) & = (E, \nu_E) (E,A) \\
	(\nu_E,\nu_A) & = (E, \nu_E) (E,\nu_A) \\
	& = (E, \nu_E) (A,\nu_A) (E,A) \\
	(\nu_E,X) & = (E, \nu_E) (E,X) \\
	(\nu_E,\nu_X) & = (E, \nu_E) (E,\nu_X) \\
	& = (E, \nu_E) (X, \nu_X)(E,X)
\end{align}
With this step we have added the last missing edges in the graph $\gamma$ and have obtained a positive definite completion of the matrix $H$.

\section{Several currencies}\label{sec.several}
\begin{figure}[hbt]
  \begin{center}
  	\includegraphics[width=10cm]{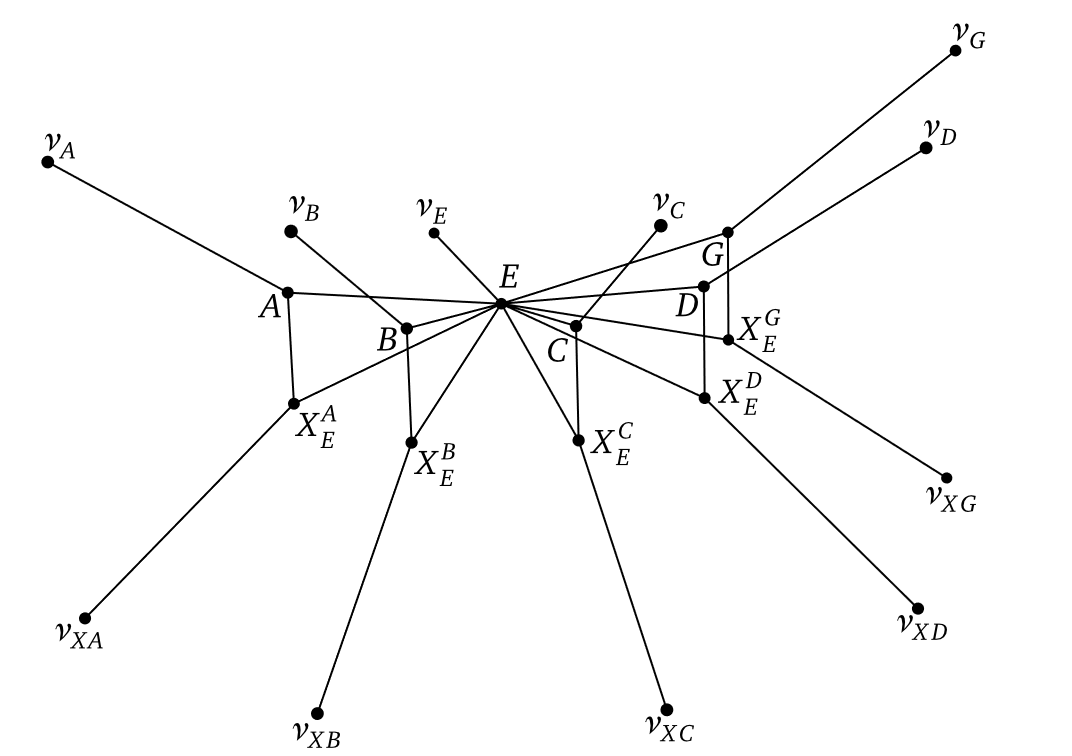}
  \end{center}
  \caption{The graph corresponding to the incomplete correlation matrix for five foreign currencies.}\label{fig.several}
\end{figure}

\begin{figure}[hbt]
  \begin{center}
  	\includegraphics[width=12cm]{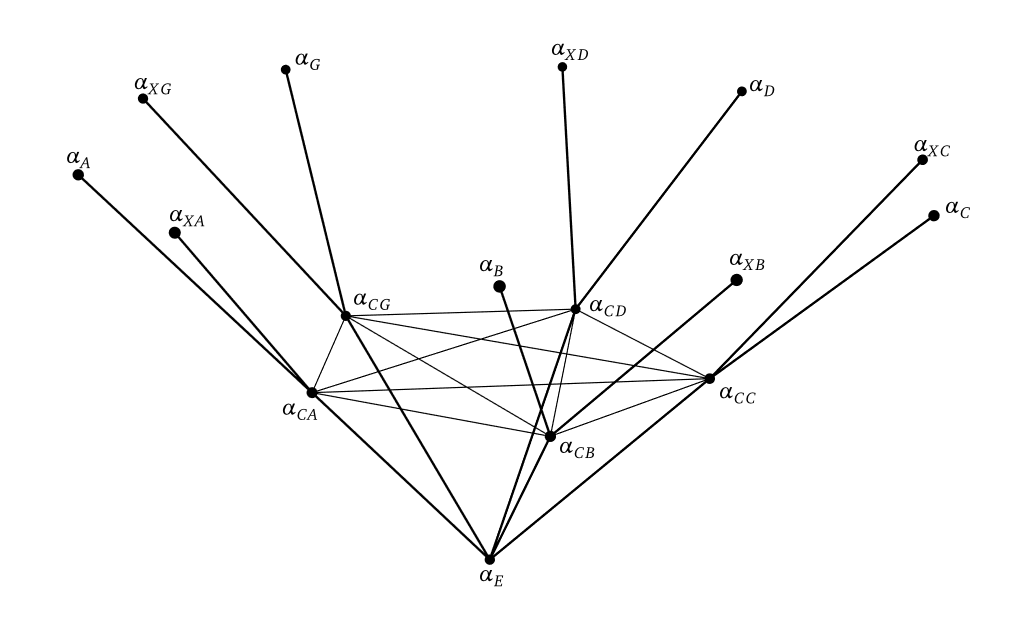}
  \end{center}
  \caption{The graph of cliques for the case of several foreign currencies. A clique tree is obtained by removing the lighter edges in the middle layer. It is again a tree of height two.}\label{fig.severalCliques}
\end{figure}

In the previous section we have just looked at one foreign currency $A$ with one exchange rate $X$. The procedure can easily be extended to an arbitrary number of foreign currencies. We will denote the foreign currencies by a capital letter:
\begin{equation}
	A, B, C, \ldots
\end{equation}
The exchange rate between the domestic currency $E$ and the foreign currency $A$ will now be denoted by 
\begin{equation}
	X_E^A.
\end{equation}
If $m_A$ is an amount in currency $A$ then $X_E^Am_A$ is the corresponding amount in currency $E$. If we are given the same correlation coefficients as in the previous section we obtain the graph in figure \ref{fig.several}. The corresponding graph of cliques is shown in figure \ref{fig.severalCliques}. The corresponding clique tree is obtained by erasing the lines that connect the cliques 
\begin{equation}
	\alpha_{CA}, \alpha_{CB}, \ldots, \alpha_{CG}
\end{equation}
on the middle layer. The intersection of any two of these cliques is given by
\begin{equation}
	\{E\} \subset \alpha_E.
\end{equation}
Since the path connecting the two cliques passes through $\alpha_E$ this tree satisfies the intersection property. 

\section{Conclusion}\label{sec.conclusion}
The increasing size of financial models has made insufficiently specified correlation matrices a common occurrence. The larger the model the fewer correlation coefficients are usually known. We thus need a rational for how to deal with these incomplete matrices as well as a robust method for implementing this rational. In this paper we have supplied both. We have argued that the correct way to complete the matrix is to choose the matrix that maximizes the entropy of the distribution described by the matrix. This is the matrix with the maximal determinant. Then we have shown how to actually construct this matrix. The procedure that we have described in section \ref{sec.procedure} is straight forward to implement and is guaranteed to produce a valid correlation matrix provided one starts with a matrix whose associated graph is chordal. Even for models with a large number of stochastic variables the procedure is both robust and fast. 

\vspace{2cm}

\section*{Disclaimer}
The views and opinions expressed here are those of the authors and do not necessarily represent the views and opinions of their employers.

\section*{Acknowledgement}
The authors would like to thank Patrick B\"uchel for his support during the creation of this work; and Nataliya Koval, and Roland Seydel for insightful discussions on the subject.

\end{document}